# Fluorescence Resonance Energy Transfer (FRET) as Biomarkers

Ripa Paul, Sudip Suklabaidya, Syed Arshad Hussain*

*Thin film and Nanoscience Laboratory, Department of Physics, Tripura University, Suryamaninagar*
*\* Corresponding author*
*Email: sa_h153@hotmail.com, sahussain@tripurauniv.in*
*Ph: +919402122510 (M), +91381 2379119 (O)*
*Fax: +913812374802 (O)*

**Abstract**

Applications of Fluorescence Resonance Energy Transfer (FRET) in biology have expanded tremendously in the last 25 years. This technique has become a staple technique in many biological and biophysical fields. FRET based chemical sensors and biosensors can play an important role as biomarkers. Development of new and effective FRET based assay for biological application is an emerging area of research needing multidisciplinary approach from biologists, chemists and physicists. Here we have discussed the use of FRET technique as biomarker.

## 1. Introduction

A biomarker may refer to as an important tool used in biological science ranging from sophisticated research to common laboratory diagnostics as indicator of normal or disordered biological state [1]. Biomarkers are objectively measured and analyzed as indication of normal biological process or pharmacological response under a specific therapeutic condition. Biomarkers may be helpful to have idea about the delivery of drugs to its target as well as to understand and predict consequent effects by monitoring certain changes of variables known to have chemical relevance.

Over the last two decades a numerous sophisticated fluorescence based spectroscopic assays have been developed in order to study biological process and to study the behaviour of biomolecules both *in vitro* (freely diffusing or immobilized molecules) and in living cells . Among these assays, FRET between two fluorophores has gained much interest due to its potential to probe the different biomolecular processes [2-7]. FRET has been widely used in all applications of fluorescence including medical diagnostics, DNA analysis, optical imaging and for various sensing properties [2-7]. Generally, fluorescence-based sensors adopt three different strategies: (a) fluorescence quenching (turn-off), (b) fluorescence enhancement (turn-on) and (c) ratiometric FRET. FRET sensors become popular tools for studying intracellular processes as well as for the understanding of several biological systems.

FRET is a physical dipole–dipole coupling between the excited state of a donor fluorophore and an acceptor chromophore that causes relaxation of the donor to a non-fluorescent ground state, which excites fluorescence in the acceptor [6-8]. In the process of FRET, initially a donor fluorophore (D) absorbs the energy due to the excitation of incident radiation and transfers the excitation energy to a nearby chromophore, the acceptor (A). The process can be expressed as follows:

$$D + h\vartheta \rightarrow D^*$$
$$D^* + A \rightarrow D + A^*$$
$$A^* \rightarrow A + h\vartheta$$

[Where $h$ is the Planck's constant and $\vartheta$ is the frequency of the incident radiation]

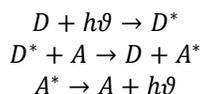

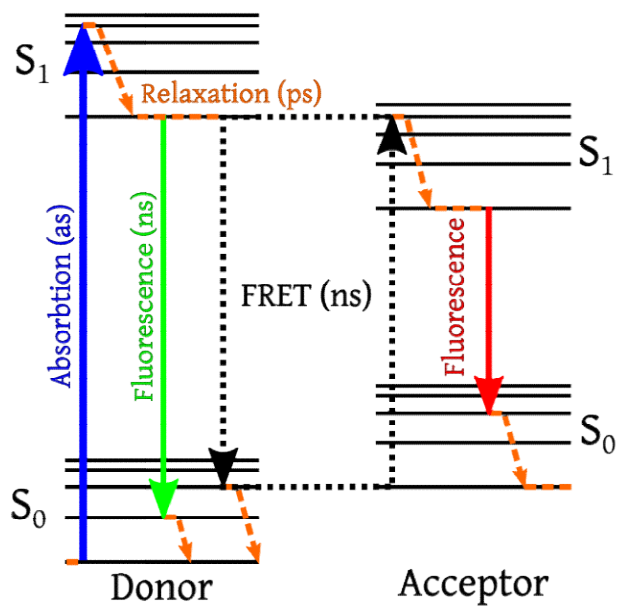

Fig. 1. Jablonski diagram illustrating the FRET process (Source: Wikipedia)

FRET is a highly distance dependent phenomenon – typically 1 to 10 nm distance between donor and acceptor fluorophore is crucial for the process to occur. There are few criteria that must be satisfied in order for FRET to occur [7-8]. These are: (a) the fluorescence spectrum of the donor molecule must overlap with the absorption or excitation spectrum of the acceptor chromophore. The degree of overlap is referred to as spectral overlap integral. (b) Two fluorophores (donor and acceptor) must be in the close proximity to one another (typically 1 to 10 nanometer). (c) The transition dipole orientations of the donor and acceptor must be approximately parallel to each other. (d) The fluorescence lifetime of the donor molecule must be of sufficient duration to allow the FRET to occur. Experimentally, FRET is measured either as the decrease in the lifetime or intensity of donor fluorescence after the addition of acceptor, or as the increase in acceptor fluorescence after the addition of donor.

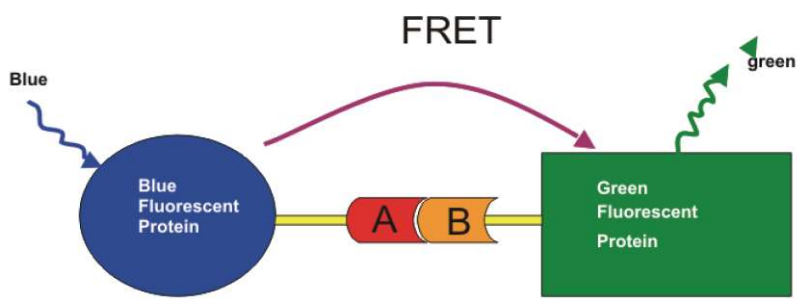

Fig. 2: Schematic representation of the interaction of two different fluorescent protein. Protein-protein interactions between proteins labeled A and B bring blue fluorescent protein and green fluorescent proteins in close enough proximity to allow for FRET to occur. In this example, excitation of blue fluorescent protein results in the emission of fluorescence by Green fluorescent protein. (Source: The application note of BioTek Instrumentswrittenby Paul Held, Laboratory Manager, Applications Department, BioTek Instruments, Inc., Winooski, VT)

T. Förster put forward an elegant theory which provided a quantitative explanation for the non-radiative energy transfer and quantitative analysis of the process [8]. Later FRET technique highlighting its application has been extensively reviewed by several researchers [6-10].

In addition to the successful applications of FRET in bulk studies, the FRET phenomenon has been successfully performed in single molecule level. Also the effective range of FRET is 1 – 10 nm, which typically within the order / dimension of most of the bio-macromolecules. This approach has been very fruitful as it has provided the biological community with new insights in large number of important biological questions e.g. which structural conformations does a protein adopt during its folding to the native state, what sort of structural changes are induced to a DNA structure when a protein binds to it, how the structure of an enzyme or protein in general is related to its functional role?

Several FRET assays for studying biochemical process have been developed. If the donor and acceptor are attached to different biomolecules (intermolecular FRET) one can study whether or not the molecules are bound to each other and thus study receptor-ligand reactions or the formation and dissociation of multi-protein complexes or DNA/RNA-protein complexes. Alternatively, if both the donor and the acceptor are attached to the same biomolecule (intramolecular FRET) one can monitor the structural changes of the molecule of interest as it performs its biological role.

## 2. Few Example

The first attempt of imaging live cellular functions using FRET was to visualize the intracellular dynamics of adenosine 3, 5- cyclic monophosphate (cAMP) by designing a probe based on cAMP-dependent protein kinase. Two flurophores suitable for FRET were labeled with the regulatory and catalytic subunits. The regulatory subunit dissociates from the catalytic subunit upon binding of cAMP resulting disappearances of FRET [11]. Multiple genetically-encoded FRET probes have been developed for use in neuronal and non-neuronal cells. These probes can be classified into several categories depending on the approach used to detect different types of biological phenomena. The interaction between proteins can be monitored by inter- molecular FRET, where one party of the protein complex is tagged by a donor and the other by an acceptor emergence of genetically encoded FRET probes. Use of FRET for a signal transduction mechanism was demonstrated by Campbell et al. [12]. They suggested that the efficiency of energy transferred from donor molecule to its receptor molecule depends on glucose binding [12]. A novel strategy using FRET-based nano-probes for the simultaneous detection of multiple mRNAs with single wavelength excitation in living cells were also reported [13].

## 3. FRET Biosensor

A FRET biosensor usually consists of a donor, a sensor domain, a ligand domain and an acceptor. The sensor domain senses the alteration in environmental signal, which induces conformational change of the sensor domain. The ligand domain recognizes the conformational change of the sensor domain, leading to change in FRET efficiency.

FRET-based biosensors take advantage of FRET, due to its high sensitivity of detection and high signal-to-noise ratio. There are two types of FRET-based sensors: those that turn "on" in the presence of analyte and the other is turn 'off' in presence of analyte [14].

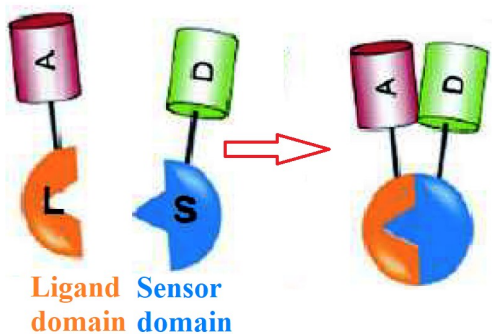

**Ligand  Sensor**
**domain domain**

Fig. 3. A nano-sensor that produced FRET signals was developed to identify Kras oncogene point mutations in ovarian tumors [reprinted from ref. 15 with permission].

FRET is a useful strategy because it enables one-pot detection of the marker, and eliminates time-consuming washing steps that would be required to separate unbound QD probes from the bound probes. Assays involving FRET have not been extensively investigated for detection of cancer biomarkers, but with FRET there is the potential to achieve lower limits of detection because of the low background perturbations. Determination of the biomarker concentration based on FRET ratio may also enable more accurate sensing.

The direct and specific detection of biomarkers is crucial as it can allow monitoring the state of a tissue or wound as well as the progression of the inflammatory process. Neutrophil Elastase (NE) plays an important role in many biological processes. It is involved in inflammatory diseases and is enriched in inflamed tissues, in wound exudate, and in the sputum of cystic fibrosis patients. FRET based biosensor assay to detect NE, were designed with a NE-specific protein, whose fluorescence features are altered upon selective cleavage by NE at physiological concentration [16].

Genetically-encoded FRET based biosensors enable us to visualize a variety of signalling events, such as protein phosphorylation and G protein activation in living cells [17]. Biosensors based on the principle of FRET have been developed to visualize the activities of the signaling molecules in living cells. Accordingly, FRET based biosensors have been used in cancer research [18-19]. Stable expression of FRET based biosensors will accelerate current trends in cancer research, that is, from cells on a plastic dish to 3-D and/or live tissues, and from biochemistry to live imaging.

## 4. Outlook

Simultaneous monitoring of multiple molecular interactions and multiplexed detection of several diagnostic biomarkers at very low concentrations have become important issues in advanced biological and chemical sensing. FRET can play an important role in this regard. Development of FRET based sensing system for practical application is a challenge, requiring an interdisciplinary outlook. Future progress of research in the area of FRET sensor is dependent upon the close collaboration of physicists, chemists, biologists, material scientists and computing specialists.

## Acknowledgements

The authors are grateful to DST, for financial support to carry out this research work through FIST – DST project ref. SR/FST/PSI-191/2014. SAH is grateful to DST, for financial support to carry out this research work through DST, Govt. of India project ref. No. EMR/2014/000234. The authors are also grateful to UGC, Govt. of India for financial support to carry out this research work through financial assistance under UGC – SAP program 2016.